# Transition-Metal-Dichalcogenide Tunable Quantum Relay Device

*Anshika Upadhyay*
*M.S. Electrical and Computer Engineering (Optics and Photonics)*
*University of Michigan, Ann Arbor, USA*

**Abstract:** One of the biggest challenges in implementation of Quantum circuits or Photonic Integrated Circuits in general is the inability to create efficient relay devices due to small decoherence time, high delays and poor interconnections that worsen the decoherence and delay problems. A feasible relay device design is proposed that can work in conjunction with various sources, be highly tunable over a reasonably large bandwidth and operate with negligible switching delays. This is achieved with the help of the Transition Metal Dichalcogenide (TMDC) $MoSe_2/WSe_2$ heterojunction and a modulating chirped strain/acoustic wave is applied to control the operating frequency and other operational characteristics of the device. The "chirping" of strain wave is meant to control the exciton transport across the device. The switching delay is proposed to reduce by exploiting the exciton dynamics so that the carrier non-equilibrium is never disturbed and therefore no delay occurs while switching on or off. $MoSe_2/WSe_2$ heterojunction is chosen because of their binding energy nearly two orders greater than GaAs quantum wells because of which they demonstrate high exciton response which is electrically tunable even at room temperature. Also, it is possible to grow TMDC films on Si based substrates using particular techniques so that their properties are intact. Such a device can find applications in wide range of components in photonic/quantum optoelectronic integrated circuits such as switches, logic gates, sensors and buffers.

## Introduction

The binding energy of an exciton (which is dependent on the spatial size of the exciton, and is negative) is what makes it different from the electrons and holes in a medium. The different exciton spatial distances (and hence, binding energy) lead to interesting physics and many exciting applications. One example is the use of excitons in photovoltaics where absorption of appropriate amount of energy leads to creation of exciton pair and conversely, excitons may annihilate under stimulating conditions and emit a photon. Exciton physics is being exploited in variety of semiconductor applications for various applications like design of photonic memory [1], laser with ultralow threshold [2], etc. Monolayer heterostructures are considered ideal exciton media as they have long-lived excitons compared to their single-layered counterparts [3]. These layers stay together by weak Vander Waals forces (and not grown on top of another). This kind of structure does away with the problems associated with lattice mismatch and allows one to choose any two materials for bandgap engineering. The combination of $MoSe_2$ – $WSe_2$ is taken here for the application. $MoSe_2$ – $WSe_2$ heterojunction forms a Staggered (type II) band alignment (figure 1) [4].

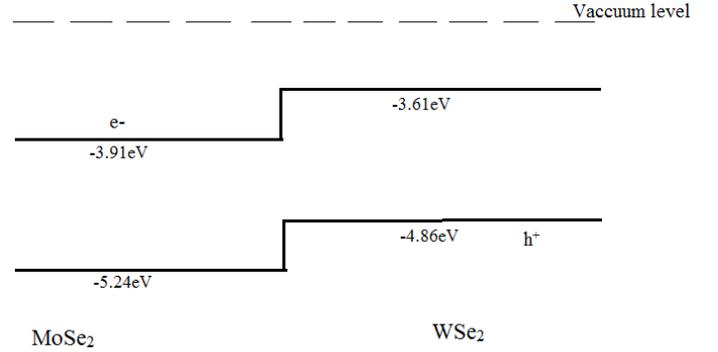

Figure 1. Band alignment of $MoSe_2$ – $WSe_2$ heterojunction

Because TMDCs have a high Coulomb binding energy of several hundred meV (much greater than typical semiconductors and nearly two orders greater than GaAs quantum wells), these semiconductors demonstrate high exciton response that is electrically tunable even at room temperature. Also, it is possible to grow TMDC films on Si based substrates using particular techniques so that their properties are intact [5].

The excitons are spatially indirect as the electron will be confined in the conduction band of $MoSe_2$ while the hole will be confined in the valence band of $WSe_2$. This is because the conduction band maximum for electron in $MoSe_2$ and valence band minimum for hole is going to occur at different spatial frequencies. This spatial indirectness is also directly measured in [3]. Such spatially indirect excitons live longer than the direct excitons and thus, such structures are very useful to exploit the exciton dynamics.

The ultimate goal of this paper is to engineer the strain signal to cause particular kind of exciton dynamics, exploiting the beneficial properties of TMDC heterojunction, to be able to use it in desired applications.

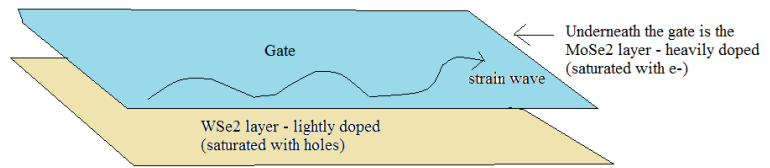

Figure 2. Proposed design of the device

## Theory, Results and Discussion

Binding energy, $B.E = \dfrac{-\hbar^2}{2 \times r_{bohr}^2 \times \mu}$ where $r_{bohr}$ is the bohr radius and $\mu$ is the reduced mass. Hamiltonian of a simple unperturbed exciton is given by $H_o = H_e + H_h + H_e$. Hamiltonian of electron, $H_e = \dfrac{-\hbar^2}{2 \times m_e^*} \dfrac{\partial^2}{\partial z^2} + V_{CB}(z)$, Hamiltonian of hole, $H_h = \dfrac{-\hbar^2}{2 \times m_h^*} \dfrac{\partial^2}{\partial z^2} - V_{VB}(z)$. Therefore, Hamiltonian of the electron-hole interaction, $H_{e-h} = \dfrac{P_\perp^2}{2 \times \mu_\perp} + \dfrac{-e^2}{4\pi\varepsilon r_{e-h}}$ where $r_{e-h}$ is the electron-hole separation and $\mu_\perp$ is the in-plane reduced mass of the electron-hole pair. $P_\perp$ is the momentum of electron-hole pair.

$$P_\perp = -\hbar^2 \left(\dfrac{\partial^2}{\partial x^2} + \dfrac{\partial^2}{\partial y^2}\right) \quad [6]$$

Adding spin-orbital coupling in each layer, $-\dfrac{e}{2m_e^{*2}c^2}\left(\dfrac{1}{r_e'}\dfrac{dV(r_e')}{dr_e'}\right)(L.\hat{S})_e + \dfrac{e}{2m_h^{*2}c^2}\left(\dfrac{1}{r_h'}\dfrac{dV(r_h')}{dr_h'}\right)(L.\hat{S})_h$ where,

$S$ is spin operator and $L$ is angular momentum. $V(r)$ is the coulombic potential (it can be taken from the Lennard Jones potential formula) and $r_e'$, $r_h'$ are respective molecular radii. $\hat{L} = \hat{R} \times \hat{P} = -i\hbar \hat{R} \times \vec{\nabla} =$
$-i\hbar\left[(z-y)\dfrac{\partial}{\partial x} + (x-z)\dfrac{\partial}{\partial y} + (y-x)\dfrac{\partial}{\partial z}\right]$.

Since the deflection in the 'z' direction is much more significant than the other two, therefore, $\hat{L}$ reduces to $\hat{L} = -i\hbar z\left(\dfrac{\partial}{\partial x} - \dfrac{\partial}{\partial y}\right)$. Therefore, the spin-orbital coupling is now given by $i\hbar f(r')\left(\dfrac{\partial}{\partial x} - \dfrac{\partial}{\partial y}\right)$ where

$f(r') = f(r_e') - f(r_h') = \dfrac{e}{2m^{*2}c^2}\dfrac{dV(r')}{dr'}$ is the net strength of the coupling. It is also the indicator of the localized electric field and can help control the interaction between the emitted photons with the electric field strength of the exerted piezoelectric/acoustic/strain signal (a polariton).

By conservation of energy, Energy (applied acoustic/strain signal) + Energy (exciton) = 0. Thus, $\hbar\Omega_{strain} = E_g - B.E$ where $E_g$ is the band-gap energy. This is important as it helps to engineer the bandgap and binding energy (or the bohr radius) for the required strain energy. Thus, the final Hamiltonian is:

$$H = \dfrac{-\hbar^2}{2m_h^*}\dfrac{\partial^2}{\partial z^2} - V_{VB}(z) + \dfrac{-\hbar^2}{2m_e^*}\dfrac{\partial^2}{\partial z^2} + V_{CB}(z) - \dfrac{\hbar^2}{2\mu_\perp}\left(\dfrac{\partial^2}{\partial x^2} + \dfrac{\partial^2}{\partial y^2}\right) - \dfrac{e^2}{4\pi\varepsilon r_{e-h}}$$
$$+ i\hbar f(r')\left(\dfrac{\partial}{\partial x} - \dfrac{\partial}{\partial y}\right) - B.E$$

where $E_g$ (band-gap energy) = $V_{CB}(z) - V_{VB}(z)$.

Taking $-V_{VB}(z) + V_{CB}(z) - \dfrac{e^2}{4\pi\varepsilon r_{e-h}} - B.E$ as -E (sign chosen for convenience), 'E' being total exciton energy.

The final time-dependent Schrodinger equation is:

$$\left[\dfrac{-\hbar^2}{2\times m_h^*}\dfrac{\partial^2}{\partial z^2} + \dfrac{-\hbar^2}{2\times m_e^*}\dfrac{\partial^2}{\partial z^2} - \dfrac{\hbar^2}{2\mu_\perp}\left(\dfrac{\partial^2}{\partial x^2} + \dfrac{\partial^2}{\partial y^2}\right) + i\hbar f(r')\left(\dfrac{\partial}{\partial x} - \dfrac{\partial}{\partial y}\right)\right.$$
$$\left. - E\right]\psi = i\hbar\dfrac{\partial \psi}{\partial t}$$

Where $\psi$ is the exciton wavefunction. Upon decoupling the equation in separate dimensions, $\psi = \psi(x)\psi(y)\psi(z)\psi(t)$, there are 4 equations to solve:

$$-i\hbar\dfrac{\partial \psi_t}{\partial t} = E_t \psi_t \qquad - (1)$$

$$-\dfrac{\hbar^2}{2\mu}\dfrac{\partial^2}{\partial z^2}\psi_z = E_z \psi_z \qquad - (2)$$

$$-\dfrac{\hbar^2}{2\mu_\perp}\dfrac{\partial^2 \psi_x}{\partial x^2} + i\hbar f(r')\dfrac{\partial \psi_x}{\partial x} = E_x \psi_x \qquad - (3)$$

$$-\dfrac{\hbar^2}{2\mu_\perp}\dfrac{\partial^2 \psi_y}{\partial y^2} - i\hbar f(r')\dfrac{\partial \psi_y}{\partial y} = E_y \psi_y \qquad - (4)$$

Equation 1 is solved with the help of the assumption that $\psi_t = c.\exp(at)$ as well as normalization and causality condition of $\psi_t$. Therefore, we obtain $E_t = \hbar\omega$ and $\psi_t = -(2\omega)^{1/3}.\exp\left(-\left(\dfrac{\omega^2}{2}\right)^{1/3} t + \dfrac{i\pi}{2}\right)$.

Equation 2 reminds us of the particle in a box scenario and thus, its solution is:

$E_z = \dfrac{n^2\pi^2\hbar^2}{2\mu d^2}$ and $\psi_z = \sqrt{\dfrac{2}{d}}\sin\left(\dfrac{n\pi z}{d}\right)$, where n = 1,2,3,…

Equation 3 is solved with the help of the assumption that $\psi_x = C_{1x}e^{k_x x} - C_{2x}e^{-k_x x}$ and the boundary conditions $\psi_x(0) = \psi'_x(0) = 0$

Upon solving, $C_{1x} = C_{2x} = C_x$.

$$\psi_x = C_x(e^{ik_x x} - e^{-ik_x x}) \text{ and } E_x^- = \frac{\hbar^2 k_x^2}{2\mu_\perp} - \hbar f(r')k_x,$$

$$E_x^+ = \frac{\hbar^2 k_x^2}{2\mu_\perp} + \hbar f(r')k_x \quad (+,- \text{ signs as superscript indicate wave propagation and reflection respectively}).$$

Equation 4 is solved similarly and the solution is obtained as:

$$\psi_y = C_y(-e^{ik_y y} + e^{-ik_y y}) \text{ and } E_y^- = \frac{\hbar^2 k_y^2}{2\mu_\perp} - \hbar f(r')k_y,$$

$$E_y^+ = \frac{\hbar^2 k_y^2}{2\mu_\perp} + \hbar f(r')k_y.$$

$C_x$ and $C_y$ are coupled or entangled with he applied strain wave and so they can be used as controls that imitate the phase change in the applied strain wave to manipulate the state changes/transitions between energy levels in the heterostructure. They will do so in order to conserve energy of the exciton pair and normalize the total exciton wavefunction.

$\psi_x$ and $\psi_y$ are taken for forward propagation only for reasons we see below.

Therefore, the final solution is:

$$E = E_x E_y E_z E_t = C_x C_y \left(\frac{\hbar^2 k_x^2}{2\mu_\perp} + \hbar k_x f(r')\right)\left(\frac{\hbar^2 k_y^2}{2\mu_\perp} + \hbar k_y f(r')\right).$$

$$\left(\frac{n^2 \pi^2 \hbar^2}{2\mu d^2}\right)\hbar\omega$$

and,

$$\psi = \psi_x \psi_y \psi_z \psi_t = C_x C_y \sqrt{\frac{2}{d}} \text{Sin}\left(\frac{n\pi z}{d}\right)(2\omega)^{1/3}.$$

$$\exp\left[\left(-t\left(\frac{\omega^2}{2}\right)^{1/3} + i\frac{\pi}{2}\right)\right]\exp\left[-i(k_x x + k_y y)\right]$$

We see that there are 4 energy splittings because of the spin-orbital coupling.

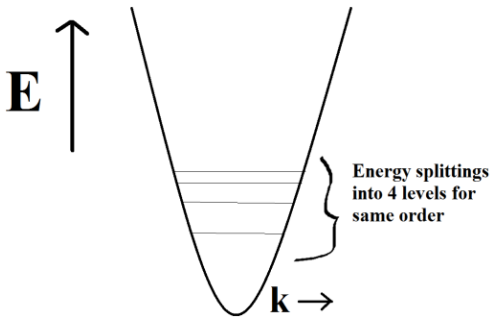

Figure 3. Energy splittings in the quantum well (heterojunction) due to spin-orbital coupling of the excitons

Clearly, if the reflection is also taken in the wavefunction then the energy splitting will be lost and also it can cause detuning of the system along with heat dissipation.

Now, E =

$$C_x C_y \left(\frac{\hbar^2 k_x^2}{2\mu_\perp} + \hbar k_x f(r')\right)\left(\frac{\hbar^2 k_y^2}{2\mu_\perp} + \hbar k_y f(r')\right)\left(\frac{n^2 \pi^2 \hbar^2}{2\mu d^2}\right)\hbar\omega$$

$$= \left|-\frac{e^2}{4\pi\varepsilon r_{e-h}} - \text{B.E} + \text{E}_g\right| = \hbar\Omega_{strain}$$

$\frac{e^2}{4\pi\varepsilon r_{e-h}}$ is basically electric field strength between the two layers multiplied by distance between the layers and unit charge.

Let B.E ~ 230meV, $E_g$ ~ 1.5eV. Let the electric field strength, $\epsilon$ = 0.317455V/Å (chosen so the dielectric behaves as semi-metal. Also, it is interesting to note that E=0 at around 0.3175V/Å) [12]. Then $\Omega_{strain}$ ~ 40GHz. Therefore, the strain frequency can be lowered by engineering

$$\left|-\frac{e^2}{4\pi\varepsilon r_{e-h}} - \text{B.E} + \text{E}_g\right| \text{ to a smaller value.}$$

Take $\mu_{MoSe2}$ (electron effective mass) ~ $0.502m_o$ and $\mu_{WSe2}$ (hole effective mass) ~ $0.341m_o$ [13]. Therefore, total effective mass ~$0.203m_o$.

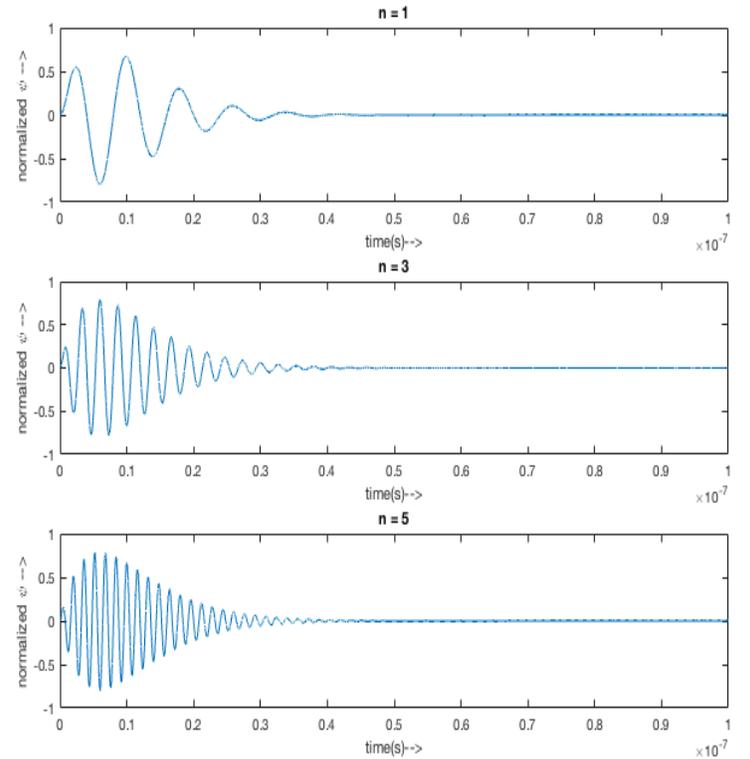

Figure 4. Normalized wavefunction of exciton.

The gate can be metal or a high-K dielectric. The distance between two layer 'd' = 4Å at which there is considerable Vander

Wal's repulsion (can take up to 7Å which is the relaxed state, i.e. no repulsion). The transition frequency is taken of the order of terahertz, it is important to keep in mind that transition energy should be less than the bandgap energy to avoid recombination. We see (from figure 4) that the decoherence time is 40-50ns which implies fast switching time. This time is not affected much by varying other parameters. Increasing 'd' reduces the magnitude of wavefunction or the total exciton probability which is expected because the dipole would break if distance is increased. Higher 'n' means faster oscillation and saturation of magnitude to a certain point. Also, longer planar dimensions will increase the exciton probability but they have to be consistent with the decoherence time otherwise the output signal will be highly attenuated and will have no meaning. Therefore, the planar dimensions are decided on the basis of a tradeoff between maximizing probability and matching with decoherence period.

Now, the constants $C_x$ and $C_y$ are entangled with the strain wave, therefore, a controlled phase modulation in the strain wave will result in spike in energy in the quantum well (figure 5).

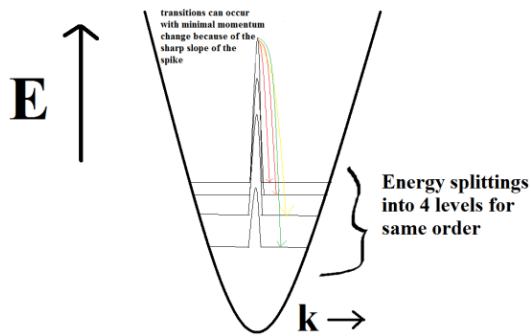

Figure 5. Spike in energy levels created due to controlled phase modulation

Also, a controlled chirp in the strain wave can cause very negligible momentum change in the exciton causing them to move forward in the quantum well without decoherence or noticeable heat dissipation (figure 6).

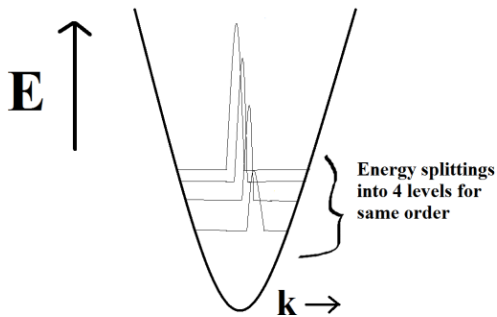

Figure 6. Dispersed energy spikes due to chirped strain wave

## Future Scope

The term $f(r')$ is also an indicator of localized electric field between the molecules. Therefore, it is interesting to study how does the bohr radius and the overall performance changes by manipulating it (in other words, how does performance differs by dealing with polaritons in the device instead of just excitons). It is also important to look into how does 'd' affect the localized electric field. Also, as the electric field modulation is easier to achieve in such a TMD heterojunction, it will be interesting to study how a quantum cascade regime changes the efficiency ([14], [15]).

It is discussed earlier that the planar dimensions and 'd' are chosen to achieve the strongest coupling between the excitons and the applied strain wave. But, it is also important to look into the width of the layers. In this paper, the study was done on a few layer thick device but it will be good to establish up to what thickness can one go in order to fit in a larger 'n' or number of energy levels without compromising with the coupling between the excitons and applied strain wave.

The gate itself needs a complete separate discussion as there will be a lot of challenges involved in achieving the right design of the device only because of gate in order to avoid problems like leakage due to quantum tunneling.

Lastly, it will be nice to look into what kind of sources can this kind of device work best with. Theoretically, the idea is to create widely tunable, all-purpose quantum relay device for quantum circuits but practically it may work better with a few sources and detectors than others.

Author Bio: Anshika Upadhyay is a Master of Science graduate from University of Michigan, Ann Arbor. She completed her degree in Electrical and Computer Engineering, specializing in Optics and Photonics.